\documentclass[%
 aps,
 amsmath,amssymb,
showkeys,
 reprint,%
]{revtex4-1}

\usepackage{graphicx}
\usepackage{dcolumn}
\usepackage{bm}
\usepackage{multirow}
\usepackage{epstopdf}

\newcommand{\braket}[2]{\left\langle#1 |  #2\right\rangle}

\newcommand{\threeJsymbol}[6]{\left( {\begin{array}{*{20}c}
   {#1} & #2 & {#3}  \\
   {#4} & {#5} & {#6} \\ \end{array}} \right)}

\begin{document}

\title{REMPI Spectroscopy of HfF}

\author{Huanqian Loh\footnote[1]{Corresponding author. Telephone: 1(303)492-7783. Fax: 1(303)492-5235. Email: loh@jilau1.colorado.edu. Postal address: JILA, University of Colorado, UCB 440, Boulder, Colorado 80309-0440, USA.} } 
 \affiliation{JILA, National Institute of Standards and Technology and University of Colorado, and Department of Physics, University of Colorado, Boulder, Colorado 80309-0440, USA}
\author{Russell P. Stutz}
\affiliation{JILA, National Institute of Standards and Technology and University of Colorado, and Department of Physics, University of Colorado, Boulder, Colorado 80309-0440, USA}
\author{Tyler S. Yahn}
\affiliation{JILA, National Institute of Standards and Technology and University of Colorado, and Department of Physics, University of Colorado, Boulder, Colorado 80309-0440, USA}
\author{Herbert Looser\footnote[2]{Permanent address: Department of Physics, University of Applied Sciences, FHNW, Windisch, 5210, Switzerland}}
\affiliation{JILA, National Institute of Standards and Technology and University of Colorado, and Department of Physics, University of Colorado, Boulder, Colorado 80309-0440, USA}
\author{Robert W. Field}
\affiliation{Department of Chemistry, Massachusetts Institute of Technology, Cambridge, Massachusetts 02139, USA}
\author{Eric A. Cornell}
\affiliation{JILA, National Institute of Standards and Technology and University of Colorado, and Department of Physics, University of Colorado, Boulder, Colorado 80309-0440, USA}

\date{\today}

\begin{abstract}
The spectrum of electronic states at 30000--33000~cm$^{-1}$ in hafnium fluoride has been studied using (1+1) resonance-enhanced multi-photon ionization (REMPI) and (1+1$'$) REMPI. Six $\Omega' = 3/2$ and ten $\Pi_{1/2}$ vibronic bands have been characterized. We report the molecular constants for these bands and estimate the electronic energies of the excited states using a correction derived from the observed isotope shifts. When either of two closely spaced $\Pi_{1/2}$ electronic states is used as an intermediate state to access autoionizing Rydberg levels, qualitatively distinct autoionization spectra are observed. The intermediate state-specificity of the autoionization spectra bodes well for the possibility of using a selected $\Pi_{1/2}$ state as an intermediate state to create ionic HfF$^+$ in various selected quantum states, an important requirement for our electron electric dipole moment (eEDM) search in HfF$^+$.
\end{abstract}

\maketitle


\section{Introduction}

The HfF$^+$ molecular ion is identified as a promising candidate for an electron electric dipole moment (eEDM) search to test fundamental symmetries and physics beyond the Standard Model \cite{LBL11, MBD06, PMI07, PMT09}. The ground state of HfF$^+$ is $X^1\Sigma^+$ \cite{PMI07, PMT09, BAB11}, whereas the low-lying ($T_e = 977$~cm$^{-1}$ \cite{BAB11, CGS12}) metastable $^3\Delta_1$ state has been predicted to offer high sensitivity for the eEDM measurement. The preparation of HfF$^+$ ions in a single ro-vibrational level of the $^3\Delta_1$ state is therefore highly desired for an eEDM search. 

Autoionization of Rydberg states of HfF, prepared using the optical-optical double resonance (OODR) technique, has been proposed as a viable way of creating HfF$^+$ ions in a single quantum state, and has been demonstrated for ion-creation in the $^1\Sigma^+$ ground state \cite{LWG11}. Since the ionization threshold of HfF lies at 59462(2)~cm$^{-1}$ \cite{BAB11, LWG11}, two ultraviolet (UV) photons of different frequencies are usually employed to access the necessary Rydberg states via the OODR technique. The final state of the autoionized product depends strongly on the intermediate state accessed by the first UV photon. Hence, there is a need to characterize and understand the neutral HfF states that lie in the vicinity of and above 30000~cm$^{-1}$, some of which may hold promise as intermediate states for HfF$^+$ ion-creation in the desired $^3\Delta_1$ state.

The HfF spectroscopy published to date includes transitions to excited states as high as 28600~cm$^{-1}$, detected by laser-induced fluorescence \cite{AHT04, neutralLIFpaper}, and resonance-enhanced multi-photon ionization (REMPI) \cite{BAB11}. This paper presents HfF spectroscopy performed in the 30000--33000~cm$^{-1}$ range, using both 1+1 REMPI and 1+1$'$ REMPI, which complements the previously published spectroscopic results.


\section{Experiment}

\begin{figure*}
\includegraphics[width=14cm]{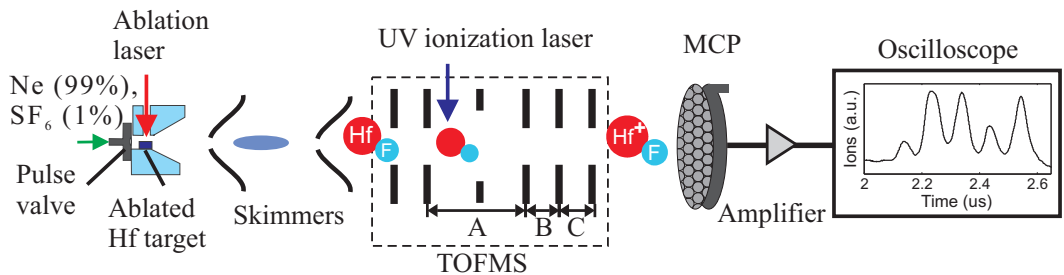}
\caption{(Color online.) Schematic of the resonance-enhanced multiphoton ionization (REMPI) experiment.}
\label{fig:tofmssetup}
\end{figure*}

\begin{figure}
\includegraphics[width=7cm]{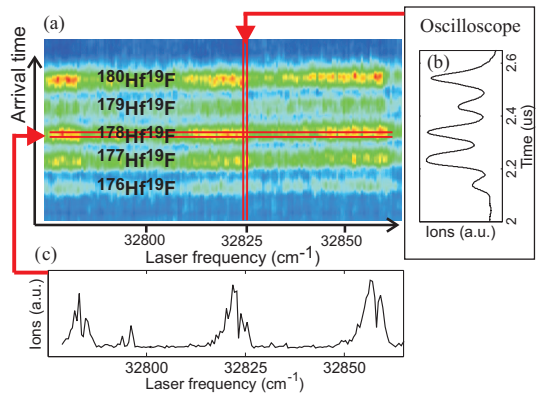}
\caption{(Color online.) (a) A pictorial summary of the oscilloscope traces taken over multiple vibronic bands. (b) A sample of the oscilloscope trace shows the capability of the time-of-flight mass spectrometry (TOFMS) apparatus to separate the spectra arising from each of the five isotopologues of HfF$^+$. (c) (1+1) REMPI spectrum of $^{178}$Hf$^{19}$F.}
\label{fig:pictorialsumm}
\end{figure}

Figure~\ref{fig:tofmssetup} shows a schematic of the experimental setup used to ionize a HfF molecular beam via (1+1) REMPI. As detailed in Ref.~\cite{LWG11}, cold ($\sim$ 10~K rotational temperature) HfF molecules are created by ablating a Hf target in the presence of 1\% SF$_6$ and 99\% Ne (690~kPa backing pressure), followed by supersonic expansion through a 800~$\mu$m diameter nozzle. The molecular beam is collimated by two 1~mm diameter skimmers separated by 29~cm before passing through a home-built time-of-flight mass spectrometer (TOFMS), which is a series of six disks arranged along a common axis \cite{Stutz09}. A UV laser, tuned such that two photons are required to ionize HfF, intersects the molecular beam perpendicularly in the TOFMS region A. The resultant HfF$^+$ ions drift under an electric field of 1~V/cm toward region B. Regions B and C of the TOFMS are operated in the Wiley-McLaren mode \cite{WM55}: when in region B, the ions are spatially focused by a small electric field of 0.25~V/cm until they reach region C, in which the ions are velocity-focused, by a transient electric field of 1~kV/cm for 500~ns, toward a microchannel plate assembly (MCP) positioned 56~cm away. The ion signal is enhanced by a transimpedance amplifier and read out by an oscilloscope. The TOFMS has a fractional mass resolution of at least 1/200, allowing us to resolve individual isotopologues of HfF$^+$ that differ by 1~amu (natural relative abundances in parentheses): $^{176}$Hf$^{19}$F (5.2\%),  $^{177}$Hf$^{19}$F (18.6\%), $^{178}$Hf$^{19}$F (27.3\%), $^{179}$Hf$^{19}$F (13.6\%) and $^{180}$Hf$^{19}$F (35.1\%). Figure~\ref{fig:pictorialsumm}(a) gives a pictorial summary of all the oscilloscope traces recorded over multiple vibronic bands, while Fig.~\ref{fig:pictorialsumm}(b) shows a sample of an oscilloscope trace for a particular wavelength of the UV laser. All five isotopologues of HfF$^+$, well-separated in arrival time on the MCP, appear as laser frequency striations of different intensities.

Most of the HfF spectra presented in this paper are recorded using (1+1) REMPI, in which the spectral resolution is limited to 0.1~cm$^{-1}$ (FWHM), the linewidth of the UV laser. The UV photoionization radiation comes from a 532~nm-pumped dye laser operating with DCM dye, frequency doubled in a KDP crystal (100-200~$\mu$J/pulse, 10~ns, 0.1~cm$^{-1}$ FWHM). Its wavelength is continuously monitored by a wavemeter calibrated against Ne spectral lines. 

Two of the HfF vibronic bands are recorded with high spectral resolution (0.003~cm$^{-1}$) using (1+1$'$) REMPI, where two co-propagating laser pulses simultaneously intersect the molecular beam at the same intersection point. The first photon (3~$\mu$J/pulse, 10~ns, 150~MHz FWHM) is the frequency doubled output of a home-built two-stage Rhodamine 101 dye cell amplifier. The dye cell amplifier is seeded by a continuous-wave ring dye laser operating with Rhodamine 610 Chloride dye. It is pumped by the second harmonic of a Nd:YAG laser. The seed laser frequency is monitored by a high-precision wavemeter that is regularly calibrated against the $^{87}$Rb $D_2$ transition. The second photon in (1+1$'$) REMPI, held at a fixed wavelength of 351.5~nm, is provided by the same dye laser as that used in the (1+1) REMPI setup, but operating with LDS722 dye. 

In this paper, transitions between energy levels of HfF are labeled as $\{ \nu_0/10^3 \}$, where $\nu_0$ is the vibronic band origin in cm$^{-1}$ \cite{BAH12}. 


\section{Results}

\begin{figure*}
\includegraphics[width=15cm]{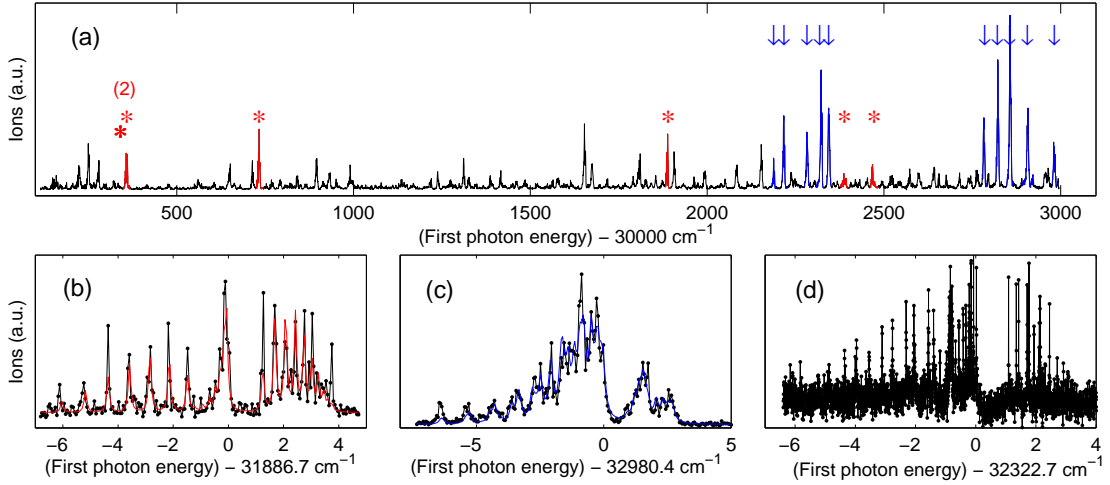}
\caption{(Color online.) (a) Coarse (1+1) REMPI scan of excited HfF states near 30000 cm$^{-1}$. The ion signal shown is the integrated ion signal for all five HfF isotopologues. Two main types of transitions have been identified in this survey scan: $\Omega' = 3/2 \leftarrow \Omega'' = 3/2$ (marked by asterisks, typical spectrum shown in (b)) and $\Pi_{1/2} \leftarrow \Omega'' = 3/2$ (marked by arrows, typical spectrum shown in (c)). (b) Detailed (1+1) REMPI scan of the $\{31.89\}$ vibronic band. The red smooth curve is a contour fit to the data shown as connected black dots. (c) Detailed (1+1) REMPI scan of the $\{32.98\}$ vibronic band. The blue smooth curve is a contour fit to the data shown as connected black dots. (d) High resolution (1+1$'$) REMPI spectrum of the $\{32.32\}$ vibronic band. Individual upper state rotational levels of opposite parity can be resolved.}
\label{fig:coarsescan}
\end{figure*}

A survey scan of HfF transitions was conducted over the frequency range 30000--33000~cm$^{-1}$ (Fig.~\ref{fig:coarsescan}(a)), which is the range encompassed by the frequency doubled output of the DCM dye. Most of the strong transitions were scanned in detail and were found to belong to either an $\Omega' = 3/2 \leftarrow \Omega'' = 3/2$ transition (Fig.~\ref{fig:coarsescan}(b)) or a $\Pi_{1/2} \leftarrow \Omega'' = 3/2$ transition (Fig.~\ref{fig:coarsescan}(c)). 

The REMPI spectra were modeled by the following Hamiltonian for the $\Omega = 3/2$ states \cite{Herzberg}:
\begin{equation} \label{eq:hamthreehalf}
E(J) = T_e + G_{\nu} + BJ(J+1) \, ,
\end{equation}
where $T_e$ is the electronic energy (difference in energy between the extrapolated minima of the $X^2\Delta_{3/2}$ and electronically excited states) and $G_{\nu}$ is the energy of the vibrational level $\nu$. For the $\Pi_{1/2}$ states, there is strong $\Lambda$-doubling and transitions from the $\Lambda$-doubled $X^2\Delta_{3/2}$ ground state into both $\Lambda$-doublets of a given rotational level can be resolved, even in the low-resolution spectra. Since no prior information is known about the parity of a given member of a $\Lambda$-doublet, the doublets are assigned `$a$'/`$b$' instead of the usual spectroscopic notation `$e$'/`$f$' \cite{BHB75}. Their energy levels can be modeled by the following polynomial \cite{FLB04}:
\begin{eqnarray} \label{eq:hamonehalf}
E^{a/b} (J) &=& T_e + G_\nu + BJ(J+1)\nonumber\\
&& \mbox{$-$/+}\,\,\, \frac{(-1)^{(J+\frac{1}{2})}}{2}(p+2q) \left( J+ \frac{1}{2} \right) \, .
\end{eqnarray}
Following Ref.~\cite{AHT04}, we assume that the $^2\Pi$ spin-orbit splitting is large enough for us to consider only the diagonal matrix element $\left\langle ^2\Pi_{1/2} \left| \mathcal{H} \right| ^2\Pi_{1/2} \right \rangle$ in our description of the $\Lambda$-doublets. Among the $\Pi_{1/2}$ states, the $^4\Pi_{1/2}$ states can borrow oscillator strength from the $^2\Pi_{1/2}$ states via spin-orbit coupling. Since transitions into the $^4\Pi_{1/2}$ upper levels would look very similar to those into $^2\Pi_{1/2}$, we have left out the superscript in the term symbol for the $\Pi_{1/2}$ levels. Based on the $\Lambda$-doubling, none of the observed spectra belong to excited states of $\Sigma$ character, which have been previously reported in the ranges 13800--14400~cm$^{-1}$ \cite{neutralLIFpaper} and 19700--20000~cm$^{-1}$ \cite{AHT04}. For both $\Pi_{1/2}$ and $\Omega' = 3/2$ types of excited electronic states, the centrifugal distortion term, $D J^2 (J+1)^2$, is neglected because the supersonic HfF molecular beam is too cold to populate rotational levels beyond $J = 21/2$, which would be required for an accurate determination of $D$.

Although all five isotopologues were observed, only the two most abundant ($^{178}$Hf$^{19}$F and $^{180}$Hf$^{19}$F) were analyzed. (In Section~\ref{sec:largeisotshift}, we extend for selected bands our analysis to the $^{177}$Hf$^{19}$F and $^{179}$Hf$^{19}$F isotopologues.) For a given transition, the two isotopologues' REMPI spectra are simultaneously fit to a contour described by a common set of fit parameters: temperature, intensity, band origin for $^{180}$Hf$^{19}$F (${^{180}}\nu_0 = {^{180}}T_e' - {^{180}}T_e'' + {^{180}}G'_{\nu'} - {^{180}}G''_{\nu''}$), band origin for $^{178}$Hf$^{19}$F (${^{178}}\nu_0$), lower state rotational constant ($B''$), excited state rotational constant ($B'$) and, where applicable, the $\Lambda$-doubling constant $(p+2q)/2$. Since the rotational constants of a given state for the two isotopologues are related to each other by the inverse ratio of their reduced masses $\mu$ (e.g., $^{180}B/^{178}B = ^{178}\mu / ^{180}\mu$), the $B$-value ratios are held fixed for the simultaneous contour fit. The resolution of our data is insufficient to detect a difference between the $(p+2q)/2$ terms for different isotopologues. 

The molecular constants for the observed HfF transitions are summarized in Table~\ref{tab:specsummary}. The numbers in parentheses indicate the $1\sigma$-standard deviation in the last digit. The standard deviation for each fit parameter was calculated from a histogram of fit parameters obtained by bootstrapping the residuals of the fit \cite{ETBootstrap}. The bootstrapping process adds the residuals, obtained after a single fit iteration, randomly to the fitted data to form a new data set, which is then re-fit against the model; this process is repeated a hundred times to get a hundred sets of fit parameters for the histogram. For the bands examined in high-resolution (1+1$'$) REMPI scans, (1+1) REMPI data for the same bands was also available for comparison. We find good agreement between the values of $\nu_0$ from the coarse (1+1) REMPI spectra and from the (1+1$'$) REMPI spectra, where the uncertainty in $\nu_0$ is the specified accuracy of the wavemeter, i.e.~0.1~cm$^{-1}$. The global systematic error of 0.1~cm$^{-1}$ is determined by how often the wavemeter is calibrated against the Ne spectral lines. For each vibronic band, because the spectra of all isotopologues are simultaneously recorded, the isotope shifts from the (1+1) REMPI scans and (1+1$'$) REMPI scans agree well when the error calculated by the bootstrap method is at least 0.01~cm$^{-1}$. Similarly, the uncertainties in $B'$ and $B''$ appear to be dominated by the bootstrap statistics when they are at least as large as 0.001~cm$^{-1}$. When the purely statistical standard deviations in the isotope shift or $\{B', B''\}$ fall below 0.01~cm$^{-1}$ and 0.001~cm$^{-1}$ respectively, the uncertainties reported in Table~\ref{tab:specsummary} have been increased to reflect an estimate of the corresponding systematic errors. We find that the purely statistical uncertainty estimated using the bootstrap method accounts well for the uncertainty in the $\Lambda$-doubling parameter.

\begin{table*} \label{tab:specsummary}
\caption{Summary of molecular constants (all in cm$^{-1}$) for observed transitions. The numbers in parentheses denote the 1$\sigma$ uncertainties, assigned as explained in the text.}
\resizebox{16cm}{!}{
\begin{tabular}{|c|c|c|c|c|c|c|c|c|c|}
\hline \multirow{2}{*}{$\Omega'$} & \multirow{2}{*}{${^{180}}\widetilde{T_e'}$} & \multirow{2}{*}{${^{180}}\nu_0 - {^{178}}\nu_0$}& \multicolumn{3}{|c|}{$^{180}$Hf$^{19}$F} & \multicolumn{3}{|c|}{$^{178}$Hf$^{19}$F} & \multirow{2}{*}{$(p+2q)/2$}\\ 
\cline{4-9} & & & $\nu_0$ & $B''$ & $B'$ & $\nu_0$ & $B''$ & $B'$  &   \\
\hline \multirow{6}{*}{3/2} & 27165(74)& -1.67(4) & 30272.97(10) & 0.282(3) & 0.249(3) & 30274.63(10) & 0.283(3) & 0.249(3) & -\\
\cline{2-10} & 27307(19)  & -1.838(10) & 30731.53(10) & 0.2864(01) & 0.2532(10) & 30733.37(10) & 0.2867(10) & 0.2535(10) & - \\
\cline{2-10} & 28137(19) & -1.150(10) & 30277.59(10) & 0.2859(10) & 0.2596(10) & 30278.74(10) & 0.2862(10) & 0.2599(10) & - \\
\cline{2-10} & 31831(19) & -0.030(10) & 31886.71(10) & 0.2832(10) & 0.2638(10) & 31886.74(10)& 0.2835(10) & 0.2641(10) & - \\
\cline{2-10} & 31833(19) & -0.335(10) & 32466.00(10) & 0.2852(10) & 0.2630(10) & 32466.34(10) & 0.2855(10) & 0.2633(12) & - \\
\cline{2-10} & 32385(56) & 0.00(3) & 32384.89(10) & 0.2810(10) & 0.2625(12) & 32384.88(10) & 0.2813(10) & 0.2628(10) & - \\

\hline \multirow{10}{*}{1/2} & 27471(19) & -2.962(12) & 32980.34(10) & 0.2823(10) & 0.2503(10) & 32983.31(10) & 0.2826(10) & 0.2506(10) & 0.0735(9) \\
\cline{2-10} & 29565(19) & -1.464(12) & 32282.12(10) & 0.2849(10) & 0.2586(10) & 32283.58(10) & 0.2852(10) & 0.2589(10) & 0.0176(6) \\
\cline{2-10} & 29682(37) & -1.43(2) & 32343.44(10) & 0.2843(10) & 0.2623(10) & 32344.87(10) & 0.2846(10) & 0.2626(10) & 0.0787(7) \\
\cline{2-10} & 29723(37) & -1.71(2) & 32905.78(10) & 0.2800(12) & 0.2499(12) & 32907.5(2) & 0.2803(12) & 0.2502(11) & 0.0159(7) \\
\cline{2-10} & 31276(37) & -0.49(2) & 32188.26(15) & 0.280(2) & 0.257(2) & 32188.75(15) & 0.281(2) & 0.257(2) & 0.057(2) \\
\cline{2-10} &  31349(37) & -0.81(2) & 32856.7(2) & 0.2841(13) & 0.2568(12) & 32857.5(2) & 0.2844(13) & 0.2571(11) & 0.054(3) \\
\cline{2-10} & 31485.1(5) & -0.4503(3) & 32322.7375(2) & 0.28364(2) & 0.25667(2) & 32323.1878(2) & 0.28395(2) & 0.25695(2) & 0.04252(3) \\
\cline{2-10} & 31629(37) & -0.62(2) & 32782.57(11) & 0.2832(10) & 0.2573(10) & 32783.2(2) & 0.2835(10) & 0.2576(10) & 0.074(2) \\
\cline{2-10} & 31690.0(4) & -0.2834(2) & 32217.15454(14) & 0.28378(2) & 0.26019(2) & 32217.43796(14) & 0.28409(2) & 0.26047(2) & 0.08537(2) \\
\cline{2-10} & 31799(19) & -0.553(14) & 32822.37(12) & 0.2814(10) & 0.2553(10) & 32822.93(10) & 0.2817(10) & 0.2556(10) & 0.0879(12) \\
\hline
\end{tabular}
}
\end{table*}


\section{Discussion}

\subsection{Lower states of observed transitions}
From previous work, the electronic ground state of HfF is $^2\Delta_{3/2}$ and for $\nu = 0$, the rotational constants are $B = $0.284 001(7)~cm$^{-1}$ and 0.283 668(6)~cm$^{-1}$ for the $^{178}$Hf$^{19}$F and $^{180}$Hf$^{19}$F isotopologues respectively \cite{AHT04}. From our own laser-induced fluorescence studies \cite{neutralLIFpaper}, we get the isotope-averaged rotational constants to be $B_{\nu=0}$ = 0.2836(4)~cm$^{-1}$, $B_{\nu=1}$ = 0.2822(10)~cm$^{-1}$ and $B_{\nu=2}$ = 0.2791(12)~cm$^{-1}$. We found \cite{neutralLIFpaper} that in our apparatus a majority of the HfF molecules are created in the ground vibronic state, although populations in multiple lower vibrational levels can be retained, athermally, in the supersonic expansion.

In this paper, all characterized transitions appear to originate from the $^2\Delta_{3/2}$ ground state of HfF. In principle, the vibrational assignment of the lower state can be inferred from a precise fit to the lower state's rotational constant and comparison against the values obtained in Ref.~\cite{neutralLIFpaper}. The vibrational quantum number of the lower state was indeed obtained from our high-resolution (1+1$'$) REMPI data (Fig.~\ref{fig:coarsescan}(d)), which shows that some bands we observe come from the ground vibronic state. However, the (1+1) REMPI data are too noisy, too broad in linewidth, and have population in too few rotational states to permit a precise enough determination of $B''$.

\subsection{A tool for understanding vibronic bands} \label{sec:sortbandtool}

The spectra of HfF in the vicinity of 30000~cm$^{-1}$ are complicated; tools are needed to sort out the spectra and to group together excited states that belong to the same electronic orgin. For most of the vibronic bands, we did not have available to us sufficiently precise determinations of both $B'$ and $B''$ to determine the vibrational numbering of upper and lower state levels. Instead, we used the isotope shift, $\Delta \nu \equiv {^{180}}\nu_0 - {^{178}}\nu_0$, to determine an estimate of the electronic energy, $\widetilde{T_e'}$, for each excited vibrational level.

Assuming that the electronic and vibrational energies are separable, we can write the vibronic band origin for the isotopologue ${^i}$Hf$^{19}$F as
\begin{equation} \label{eq:nu0def}
{^{(i)}}\nu_0 = {^{(i)}}T_e' - {^{(i)}}T_e'' + {^{(i)}}G_{\nu'}' - {^{(i)}}G_{\nu''}'' \, ,
\end{equation}
and, assuming an approximately harmonic potential, we can write
\begin{equation} \label{eq:harmonicG}
{^{180}}G_{\nu'}' - {^{(i)}}G_{\nu'}' = \left(\frac{\sqrt{{^{(i)}}\mu} - \sqrt{{^{180}}\mu}}{\sqrt{{^{(i)}}\mu}}\right) {^{180}}G_{\nu'}' \, ,
\end{equation}
where $^{(i)}\mu$ is the reduced mass for isotope $(i)$. A similar relation applies for $G_{\nu''}''$, thus if we define ${^{180}}T_e'' \equiv 0$, we combine Eqs.~(\ref{eq:nu0def}) and (\ref{eq:harmonicG}) to give an estimated value of the electronic energy, ${^{180}}\widetilde {T_e}$:
\begin{eqnarray}
{^{180}}\widetilde{T_e} &\equiv& {^{180}}T_e' - {^{180}}T_e'' \nonumber \\
&=& {^{180}}\nu_0 + \eta\left[ \Delta\nu \right. \nonumber \\
&& \left.- \left({^{180}}T_e' - {^{178}}T_e'\right) + \left({^{180}}T_e'' - {^{178}}T_e''\right) \right] \,  ,
\end{eqnarray}
where $\eta \equiv \sqrt{{^{178}}\mu}/(\sqrt{{^{180}}\mu}-\sqrt{{^{178}}\mu}) \approx 1861$. Then, assuming that the electronic contribution to the isotope shift is much smaller than the vibrational contribution \cite{Herzberg},
\begin{equation} \label{eq:t0elmeanmu}
{^{180}}\widetilde{T_e'} = {^{180}}\nu_0 + \eta \Delta\nu \, .
\end{equation}
Even in the case that the electronic contribution to the isotope shift (${^{180}}T_e' - {^{178}}T_e' - {^{180}}T_e'' + {^{178}}T_e''$) is \textit{not} much smaller than the vibrational contribution (${^{180}}G'_{\nu'} - {^{178}}G'_{\nu'} - {^{180}}G''_{\nu''} + {^{178}}G''_{\nu''}$), two bands that share common upper and lower electronic states should get, from Eq.~\ref{eq:t0elmeanmu}, very similar values for ${^{180}}\widetilde{T_e'}$, even if $\nu''$ and $\nu'$ are different for the two bands. Again, this depends on the assumption of harmonic potentials and of separable electronic and vibrational contributions to the band energy.

\begin{figure}
\includegraphics{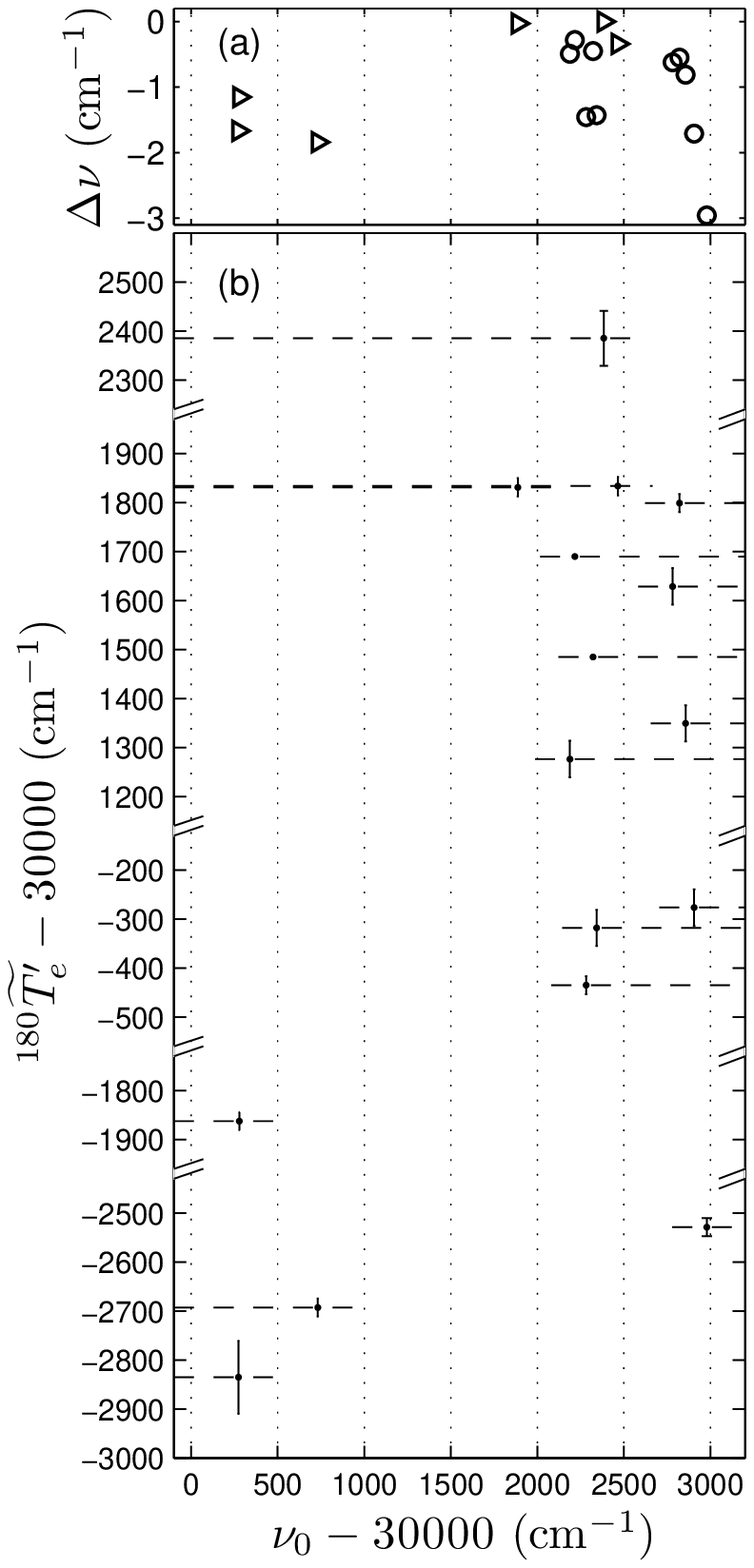}
\caption{(a) Isotope shifts (triangles denoting $\Omega' = 3/2 \leftarrow X^2\Delta_{3/2}$ transitions and circles denoting $\Pi_{1/2} \leftarrow X^2\Delta_{3/2}$ transitions) and (b) calculated electronic energies for the $\Omega' = 3/2 \leftarrow X^2\Delta_{3/2}$ transitions (dashed lines terminating on the left axis) and $\Pi_{1/2} \leftarrow X^2\Delta_{3/2}$ transitions (dashed lines terminating on the right axis). The spread in ${^{180}}\widetilde{T'_e}$ values indicates at least four distinct $\Omega' = 3/2$ electronic states and at least six distinct $\Pi_{1/2}$ electronic states.}
\label{fig:isotshift1p5n0p5}
\end{figure}

Figure~\ref{fig:isotshift1p5n0p5} is a graphical summary of the observed $\Omega' = 3/2 \leftarrow X^2\Delta_{3/2}$ bands and $\Pi_{1/2} \leftarrow X^2\Delta_{3/2}$ bands in HfF. Part (a) of the figure shows the isotope shifts, where an isotope shift is estimated to be about -0.3~cm$^{-1}$ for a change in vibrational quanta of $\nu' - \nu''= +1$. Some isotope shifts of anomalously large magnitude (1.5~cm$^{-1}$ to 3~cm$^{-1}$) are further discussed in Section~\ref{sec:largeisotshift}. Part (b) shows the estimated electronic energy versus band origin, in which transitions that share the same pair of lower and upper electronic states are expected to line up horizontally. In Fig.~\ref{fig:isotshift1p5n0p5}(b) we see two bands $\{31.89\}$ and $\{32.47\}$ that, while their band origins are separated by 579~cm$^{-1}$, have ${^{180}}\widetilde{T_e'}$ values that differ by only 13~cm$^{-1}$ (identical within measurement uncertainty). It is likely that these two bands share a common electronic transition with ${^{180}}\widetilde{T_e'} \approx 31840$~cm$^{-1}$, as indicated by the y-axis. We note that 579~cm$^{-1}$ would be a reasonable excited state vibrational spacing. We can see, from the spacing of data points in the $y$-direction in Fig.~\ref{fig:isotshift1p5n0p5}(b), that there are at least four distinct $\Omega'=3/2$ electronic levels: near 27200, 28150, 31850 and 32400~cm$^{-1}$. As a caveat, we note that several of the bands have unreasonably large isotope shifts. We discuss possible causes for anomalous isotope shifts in Section~\ref{sec:largeisotshift} below.

For the $\Pi_{1/2} \leftarrow X^2\Delta_{3/2}$ transitions, from Fig.~\ref{fig:isotshift1p5n0p5}(b) we can see immediately that there are multiple distinct $\Pi_{1/2}$ electronic levels: one with ${^{180}}\widetilde{T_e'} \approx 27500$~cm$^{-1}$, at least one with ${^{180}}\widetilde{T_e'} \approx 29600$~cm$^{-1}$, and multiple levels with ${^{180}}\widetilde{T_e'}$ in the range $31500 \pm  300$~cm$^{-1}$. One cannot say for certain that the three bands with ${^{180}}\widetilde{T_e'} \approx 29600$~cm$^{-1}$ share a common electronic transition. With reference to Table~I, we see the $\{32.34\}$ band has a $\Lambda$-doubling constant quite distinct from that of the $\{32.91\}$ band, which suggests either two distinct electronic transitions or, more likely, a local perturbation of one of the bands. For the cluster of six $\Omega'=1/2$ bands with ${^{180}}\widetilde{T_e'}$ between 31200 and 31800~cm$^{-1}$, the spread in ${^{180}}\widetilde{T_e'}$ is too large for the bands to share a common electronic transition. Moreover, the respective rotational constants for the bands $\{32.22\}$ and $\{32.32\}$ are determined so precisely that we can say with some confidence that these two bands share a common lower level vibrational level ($\nu''=0$), but belong to distinct upper vibrational levels. Since the values of $\nu_0$ are too close together to make it likely that the upper levels are adjacent vibrational levels of the same electronic state, it seems quite likely that the two bands terminate in distinct electronic upper levels. For the bands $\{32.19\}$ and $\{32.86\}$, the vibronic band origins differ by 668.41(14)~cm$^{-1}$, which is within 2$\sigma$ of the 1--0 vibrational energy splitting of the $X^2\Delta_{3/2}$ ground state \cite{neutralLIFpaper}. Further, both bands fit to the same $B'$ whereas one band fits to a value of $B''$ that is consistently smaller than that of the other band. This strongly suggests that the bands $\{32.19\}$ and $\{32.86\}$ arise from the $\nu'' = 1$ and $\nu'' = 0$ ground vibrational levels, respectively.

\subsection{Rotational line strengths} \label{sec:rotlinestrengths}

In the (1+1) REMPI experiments, the transition excited by the second photon is typically left unsaturated, so the observed REMPI line strengths are actually a convolution of the line strengths for both the first and second step in REMPI. We have observed in other work \cite{LWG11} that the above-threshold ionization spectrum is highly structured. This structure corresponds to $\nu_{\mbox{\small{Ryd}}} > 0$ levels that autoionize into $\nu^+ = 0$ vibrational levels of HfF$^+$. Qualitatively, the H\"{o}nl-London expressions for Hund's case (a) molecules work well to describe the observed rotational line intensities. This is the case for all observed bands except two $\Omega'=3/2 \leftarrow X{^2}\Delta_{3/2}$ bands: $\{30.73\}$ and $\{32.47\}$. 

To fit the line intensities in a given vibronic band, we included terms that described possible interference effects between parallel ($\Omega' - \Omega'' = 0$) and perpendicular ($\Omega' - \Omega'' = \pm 1$) transitions \cite{FLB04}. For example, a nominal $\Omega' = 3/2$ (for clarity, also denoted here as $\Omega'_N$) excited state could also possess some $\Omega' = 1/2$ (= $\Omega'_N - 1)$ and/or $\Omega' = 5/2$ (= $\Omega'_N + 1)$ character from mixing (via the Hamiltonian term -$B\textbf{J}^{\mp}\textbf{L}^{\pm}$) with nearby ``dark'' states. When the excited state is no longer a good Hund's case (a) state due to these admixtures, transitions between the $^2\Delta_{3/2}$ ground state and all three admixed $\Omega'$ components are allowed. The line strength $S(J', \Omega'_N; J'' \Omega'')$ for a given rotational line is given by:
\begin{eqnarray} \label{eq:linestrengthinterf}
&& S(J', \Omega'_N; J'' \Omega'') \propto \nonumber \\
&& \left|\mu_\alpha \threeJsymbol{J'}{1}{J''}{-\Omega'_N}{(\Omega'_N-\Omega'')}{\Omega''} \right. \nonumber \\
&& + \mu_\beta \threeJsymbol{J'}{1}{J''}{-(\Omega'_N+1)}{(\Omega'_N+1-\Omega'')}{\Omega''} \nonumber \\
&& \left. + \mu_\gamma \threeJsymbol{J'}{1}{J''}{-(\Omega'_N-1)}{(\Omega'_N-1-\Omega'')}{\Omega''}\right|^2 \, ,
\end{eqnarray}
where $\mu_\alpha, \mu_\beta$, and $\mu_\gamma$ refer to the transition dipole matrix elements between the ground state and the admixed excited states (of $\Omega'_N$, $\Omega'_N + 1$, and $\Omega'_N - 1$ character, correspondingly). Note that, since the $\mu_\alpha, \mu_\beta, \mu_\gamma$ amplitudes are summed and then squared, interference effects are present. These matrix elements are allowed to vary in the contour fits. If the second and third terms are ignored, the line strength expression reduces to the normal H\"{o}nl-London factor \cite{Zare}. 

\begin{figure*}
\includegraphics[width=15cm]{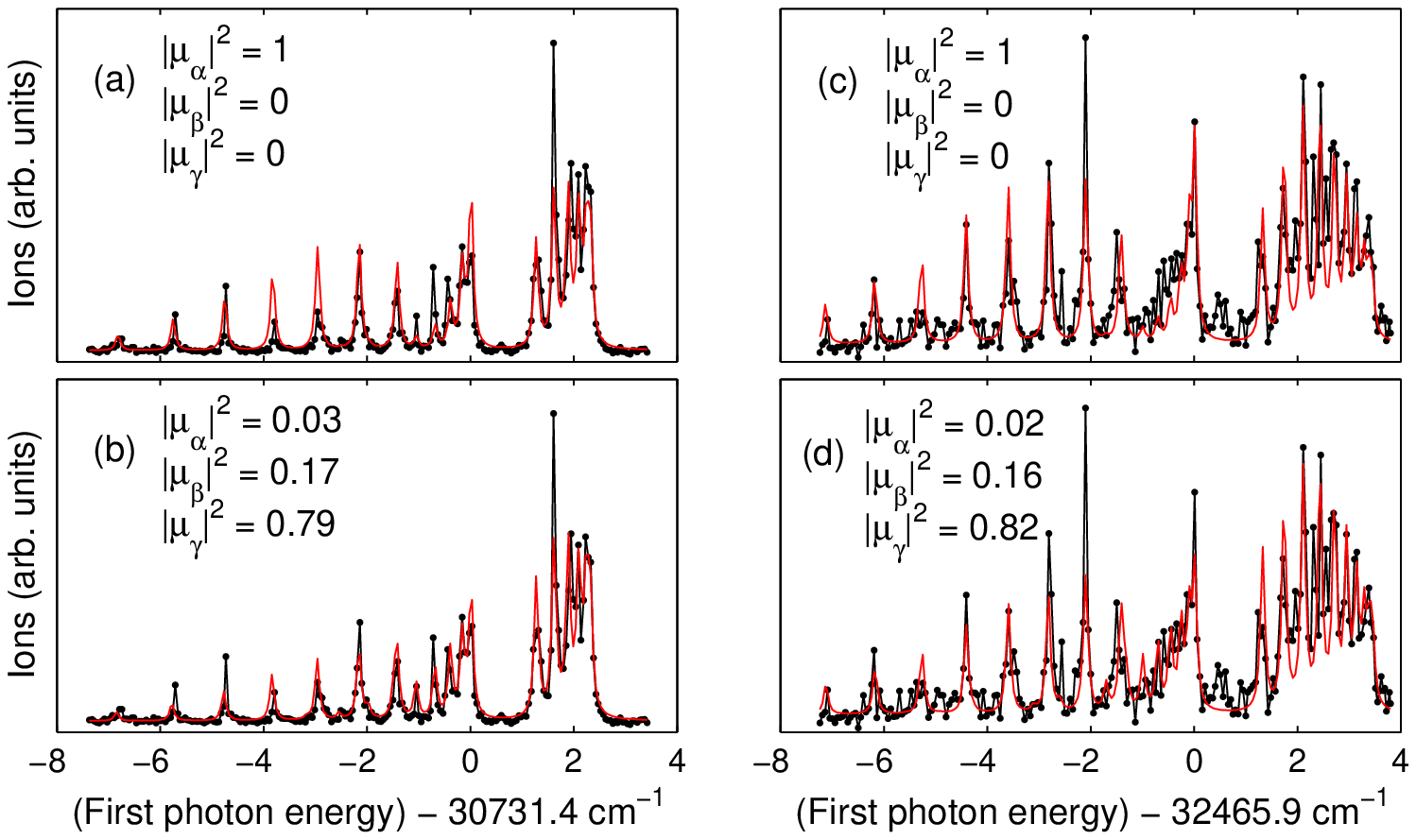}
\caption{(Color online.) Contour fits (red) to the $\Omega' = 3/2 \leftarrow X^2\Delta_{3/2}$ $\{30.73\}$ and $\{32.47\}$ bands (data shown as connected black dots). In (a, c), the contour fits are constrained to use H\"{o}nl-London expressions for the rotational line strengths, which are shown here to be inadequate for a good description of the observed line intensities, especially for those in the Q branch. In (b, d), the contour fits are allowed to take into account possible interference effects between parallel and perpendicular transition moments. In both cases, $\mu_\gamma$ dominates over the other transition dipole matrix elements, indicating that the nominal $\Omega' = 3/2$ upper states in both cases appear to possess an admixture from nearby $\Omega' = 1/2$ states. }
\label{fig:acfits}
\end{figure*}

For the $\Pi_{1/2}$ excited states, the $\Delta\Omega = \pm 1, 0$ transition selection rule requires the last term to be zero. Between the two remaining transition dipole matrix elements, $\mu_\alpha$ tends to dominate for all of the observed $\Pi_{1/2} \leftarrow X^2\Delta_{3/2}$ bands. Similarly, $\mu_\alpha$ tends to be the dominant dipole matrix element for all of the $\Omega' = 3/2 \leftarrow X^2\Delta_{3/2}$ bands except for two, $\{30.28\}$ and $\{32.47\}$. Figure~\ref{fig:acfits} compares the contour fits performed using the H\"{o}nl-London expressions versus Eq.~(\ref{eq:linestrengthinterf}) to describe rotational line intensities for each of the anomalous bands. In both cases, the incorporation of interference effects in Eq.~(\ref{eq:linestrengthinterf}) gives rise to a better fit. In particular, the contour fits indicate an admixture from a nearby $\Omega' = 1/2$ state (where $\mu_\gamma$ is dominant), which suggests that the observed upper state is of nominal $^2\Pi_{3/2}$ character, the transition ``brightness'' from the ground state has been lent to it from a nearby $\Pi_{1/2}$ state. In fact, the rotational line strengths imply that at least one of the two bands ($\{32.47\}$) is located very close to a $\Pi_{1/2} \leftarrow X{^2}\Delta_{3/2}$ band, as seen in Fig.~\ref{fig:isotshift1p5n0p5}. The observed admixture of electronic states is the reason for using Hund's case (c) notation to specify the $\Omega' = 3/2$ excited states.

\subsection{Large isotope shifts}\label{sec:largeisotshift}

\begin{figure}
\includegraphics{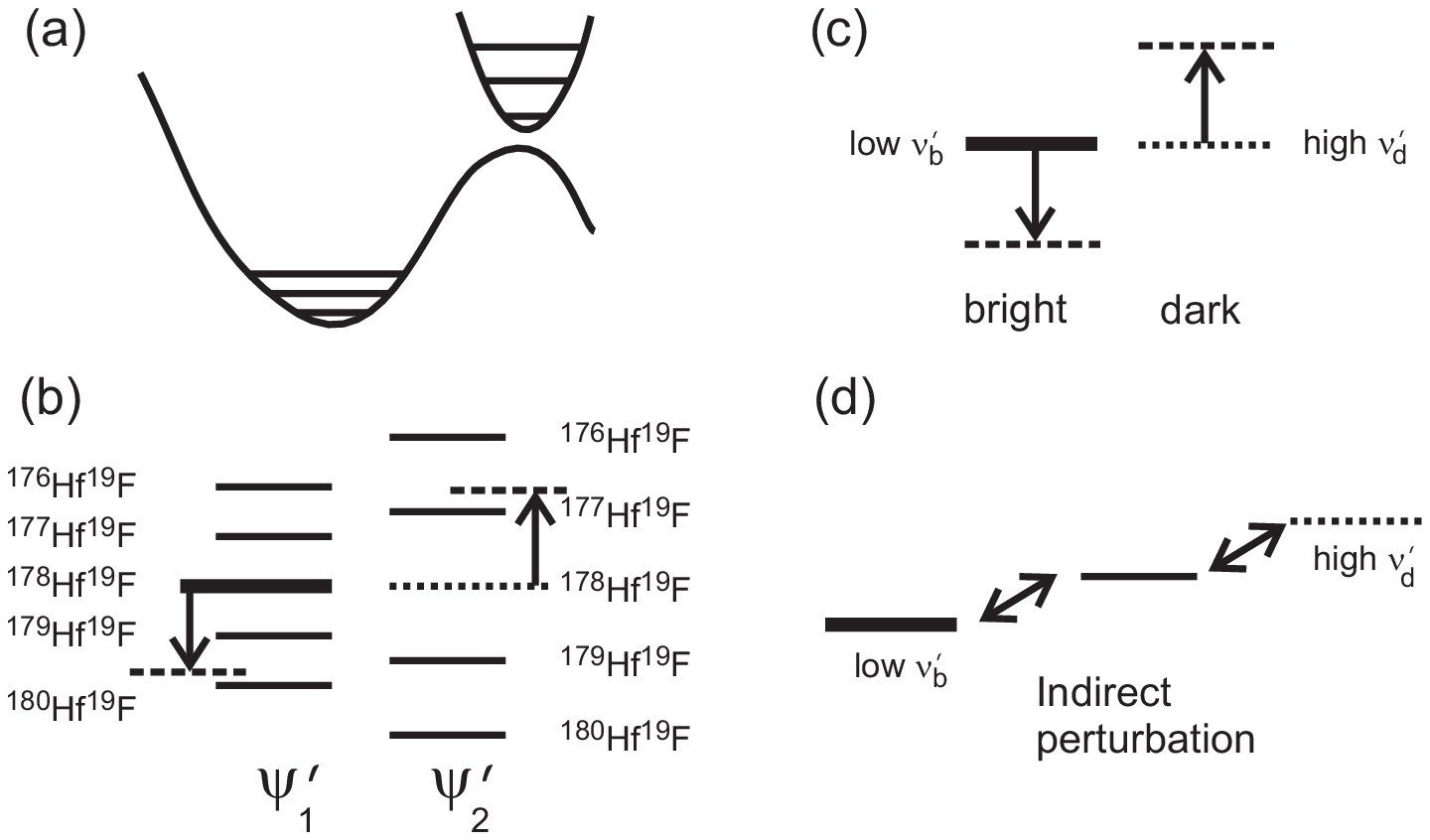}
\caption{Four hypotheses for the observation of large isotope shifts. (a) Avoided crossing of potential energy curves, giving rise to anomalously large vibrational spacings. (b) Isotope-specific accidental degeneracies between two mutually perturbing excited states, $\psi_1'$ and $\psi_2'$. In this picture, the isotope splittings would not be linear in $\sqrt{1/{^{(i)}}\mu}$. (c) Local perturbation between a high-$\nu'_d$ level of a ``dark'' electronic state and a low-$\nu'_b$ level of a ``bright'' electronic state. The higher $\nu'_d$ level becomes observable by borrowing brightness from the low-$\nu'_b$ level, but it has the normal isotope shift of a high-$\nu'_d$ level. (d) Indirect coupling between the high-$\nu'_d$ level and low-$\nu'_b$ level.}
\label{fig:largeisotshiftscheme}
\end{figure}

As shown in part (a) of Fig.~\ref{fig:isotshift1p5n0p5}, several bands were observed to have isotope shifts more negative than -1.6~cm$^{-1}$, which would correspond to the upper state of the transition having vibrational energy that exceeds the lower state vibrational energy by more than 3000~cm$^{-1}$. If we assume a typical upper-state vibrational constant of $\sim$ 600~cm$^{-1}$, a 3000~cm$^{-1}$ vibrational change implied a change in vibrational quanta, $\Delta \nu \sim 5$. Deriving bond lengths from observed rotational constants, one calculates extremely small Franck-Condon factors for such transitions. This poses a mystery: how were we able to excite these transitions? One possible explanation could be that interactions between potential curves, say an anticrossing of some sort, generate a few levels described by anomalously large vibrational constants, as depicted in Fig.~\ref{fig:largeisotshiftscheme}(a). An upper-state vibrational constant as large as 1000~cm$^{-1}$, for instance, could yield an acceptably large Franck-Condon overlap for the corresponding $\Delta \nu$ = 3 transition. But it seems unlikely there would be multiple electronic states with such exotically large curvatures in their potential curves. Moreover, in Fig.~\ref{fig:isotshift1p5n0p5}, the $x$-axis spacing between some pairs of points that are clumped along the $y$-axis suggests that at least some of the large-isotope-shift bands have relatively modest upper-state vibrational spacing, perhaps 450--600~cm$^{-1}$.

\begin{figure*}
\includegraphics[width=15cm]{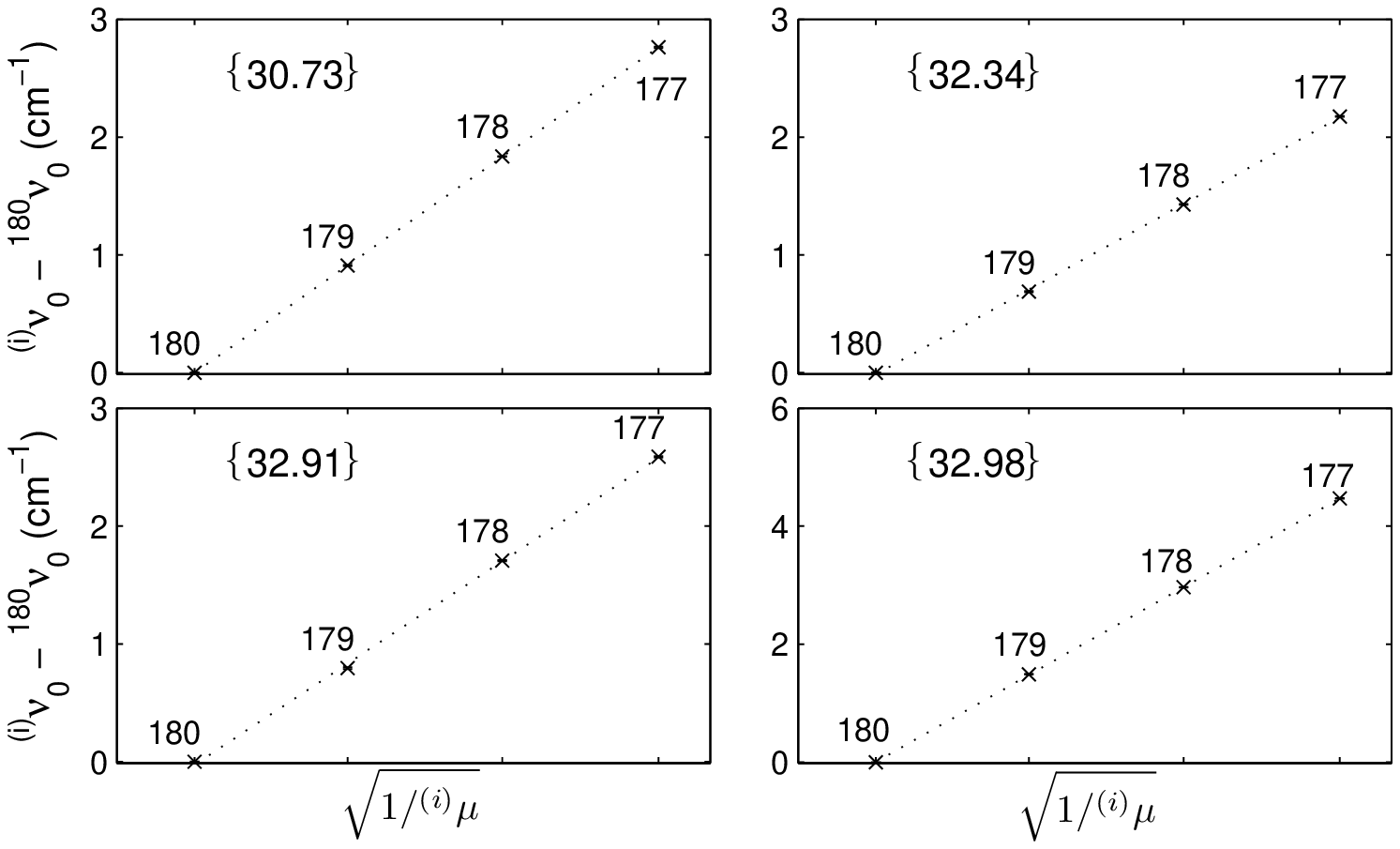}
\caption{For the vibronic bands $\{30.73\}, \{32.34\}, \{32.91\}$, and $\{32.98\}$, the isotope shifts are found to vary linearly with the inverse square root of the isotopologue reduced masses, which suggests that anomalously large isotope shifts are indeed due to a large $\Delta \nu$ and not to isotope-specific perturbations.}
\label{fig:isotsplitting}
\end{figure*}

As an alternative hypothesis to the picture of transitions involving large changes in the number of vibrational quanta, the anomalous isotope shifts could arise from isotope-specific accidental degeneracies between two mutually perturbing excited states, each of which have, in the absence of perturbation, smaller isotope shifts (Fig.~\ref{fig:largeisotshiftscheme}(b)). In this case, the perturbing state would not only cause the $^{178}$Hf$^{19}$F and $^{180}$Hf$^{19}$F states to split apart from each other, but also cause all of the other isotopologues to follow an irregular splitting that is not linear with $\sqrt{1/{^{(i)}}\mu}$. Figure~\ref{fig:isotsplitting} displays the isotope splitting between the four most abundant isotopologues for several bands with large isotope shifts. The isotopologues are found to follow a linear energy spacing relative to the inverse square root of their reduced masses, which is what we expect in the \textit{absence} of such an isotopologue-specific perturbation and where the observed isotope shift is determined primarily by $\sqrt{1/{^{(i)}}\mu}$ vibrational energy shifts.

The high visibility of bands that had large isotope shifts could also be due to spin-orbit interactions between vibrational levels that belong to different excited states, as shown in Fig.~\ref{fig:largeisotshiftscheme}(c). The observed excited state of a high vibrational quantum number, $\nu'_d$, could have been an initially ``dark'' electronic state that became observable by spin-orbit interaction with a ``bright'' electronic state of much lower vibrational quantum number, $\nu'_b$. This would provide the latter with decent Franck-Condon overlap with the lower state. Such a spin-orbit interaction would be proportional to the vibrational overlap matrix element, $\braket{\nu'_d}{\nu'_b}$ \cite{FLB04}, which is estimated to be large only for $\nu'_d \approx \nu'_b \pm 1$ for all the observed excited states in HfF.

One plausible explanation for the large isotope shifts is that there may be a third state that induces an indirect interaction between the observed state and another ``bright'' state, where the ``bright'' state is of a significantly lower vibrational quantum number and has large Franck-Condon overlap with the lower state (Fig.~\ref{fig:largeisotshiftscheme}(d)). It is likely that there are other plausible explanations for the observation of such anomalously large isotope shifts, and we invite the interested spectroscopist to offer his or her ideas.

\subsection{Intermediate states for OODR autoionization}

\begin{figure*}
\includegraphics[width=15cm]{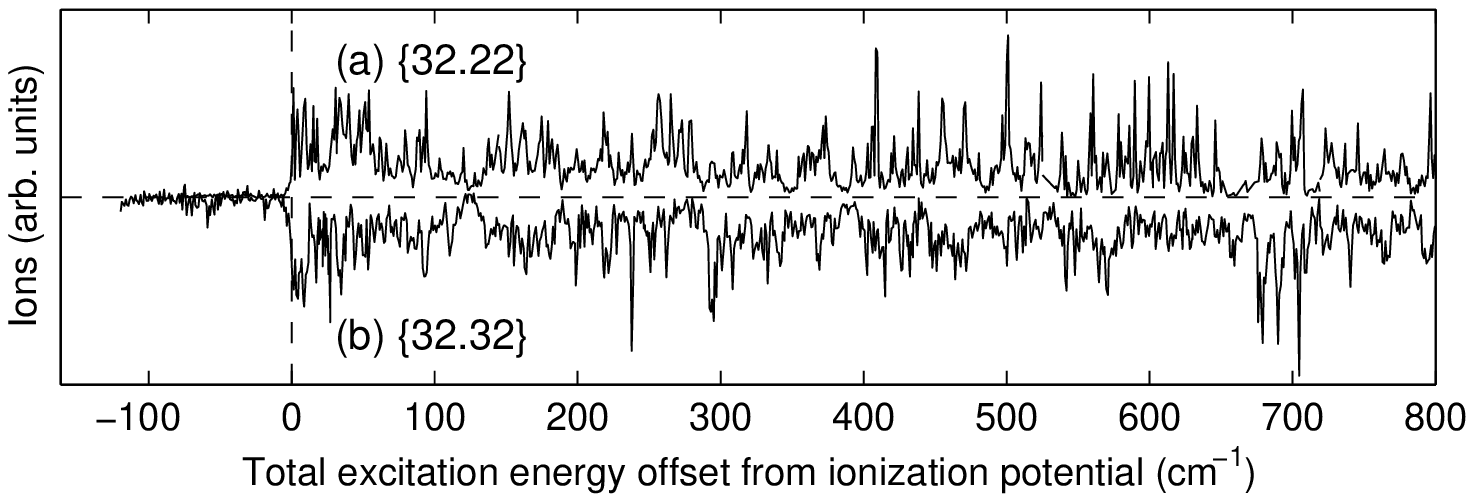}
\caption{OODR autoionization spectra measured by tuning the first photon to access a particular excited rotational state ($J' = 1/2 (a)$) using the vibronic bands (a) $\{32.22\}$ and (b) $\{32.32\}$ (ion signal inverted for clarity), then scanning the second photon to map out the spectrum of autoionizing Rydberg states accessible from that intermediate state. Each case gives a unique set of autoionization resonances, suggesting that the ionic core of the autoionizing Rydberg state differs between the two cases.} 
\label{fig:ionthresholds}
\end{figure*}

Among the six $\Pi_{1/2} \leftarrow X^2\Delta_{3/2}$ bands with ${^{180}}\widetilde{T_e'}$ in the range 31200--31900~cm$^{-1}$, two of the bands ($\{32.22\}$ and $\{32.32\}$) have been scanned at high resolution, using the (1+1$'$) REMPI technique. Their distinct $\Lambda$-doubling constants, coupled with the close proximity of the electronic energy levels, indicate that they must belong to electronically distinct states in order to exhibit such small repulsion. Their dominant electronic configurations must differ by at least two spin-orbitals, e.g.~$sd\delta(^3\Delta_1) n\ell \lambda$ versus $s^2(^1\Sigma^+)n' \ell' \lambda'$, where $sd\delta(^3\Delta_1)$ or $s^2(^1\Sigma^+)$ refers to the core configurations, and $n\ell \lambda$ or $n' \ell' \lambda'$ refers to an additional electron in a more highly excited orbital. These two configurationally distinct states could potentially be used as intermediate states to access different states of HfF$^+$ when performing OODR autoionization, which is relevant to our goal of preferential population of the metastable $^3\Delta_1$ state rather than $X^1\Sigma^+$ for the eEDM experiment. These two vibronic states have indeed been observed to yield very different autoionization spectra, as shown in Fig.~\ref{fig:ionthresholds}. We are in the process of characterizing the electronic states of the ions so produced.

\section{Conclusions}
A plethora of HfF bands in the 30000--33000~cm$^{-1}$ region have been observed using (1+1)REMPI and (1+1$'$) REMPI. We have characterized six $\Omega' = 3/2 \leftarrow X^2\Delta_{3/2}$ and ten $\Pi_{1/2} \leftarrow X^2\Delta_{3/2}$ vibronic bands. To sort out the spectra, we used the isotope shift for a given band to determine the electronic energy, ${^{180}}\widetilde{T_e'}$, for the upper electronic state. This method of grouping bands only works for bands where the potential energy curves are fairly well-approximated by a harmonic potential. Two bands exhibit interference effects between parallel and perpendicular transition moments through their rotational line strengths. Several bands with anomalously large isotope shifts had intensities much larger than predicted based on the expected small Franck-Condon factors for transitions from the low-$\nu''$ lower state. Among the six $\Pi_{1/2} \leftarrow X^2\Delta_{3/2}$ bands with electronic energy offsets crowded in the vicinity of 31300--31800~cm$^{-1}$, there are at least two electronically distinct states. When such configurationally distinct states are used as intermediate states in the OODR preparation of Rydberg states, these states provide access to at least two possible routes for creating distinct HfF$^+$ electronic states after autoionization decay, which will be important for the selective formation of ionic HfF$^+$ in the desired $^3\Delta_1$ quantum state for the eEDM experiment.

\section{Acknowledgments} 
We thank Chris Greene, Jia Wang and Matt Grau for helpful discussions. This work was funded by the National Science Foundation and the Marsico Research Chair. R.~W.~Field acknowledges support from NSF grant number 1058709. H.~Loh acknowledges support from A*STAR (Singapore).


%


\end{document}